\documentclass[%
 reprint,
 amsmath,amssymb,
 aps,
]{revtex4-2}

\usepackage{euler}
\usepackage{eucal}
\usepackage{graphicx} 
\usepackage{verbatim} 
\usepackage{color} 
\usepackage{subfigure} 
\usepackage{hyperref} 
\usepackage{soul}
\usepackage{hyperref}
\usepackage{tabularx}
\usepackage{amsmath,blkarray}
\usepackage{float}
\usepackage{amssymb}
\usepackage{mathtools}
\usepackage{setspace}
\usepackage{amsmath}
\usepackage{amsfonts}
\usepackage{bm}
\usepackage{colortbl}
\usepackage{outlines}
\usepackage{xcolor,cancel}
\usepackage{subfigure}
\usepackage[utf8]{inputenc}
\usepackage[T1]{fontenc}
\usepackage{mathptmx}
\usepackage{dcolumn}
\usepackage{float}
\usepackage{tcolorbox}
\usepackage{graphicx}
\usepackage{multirow}

\usepackage{graphicx}
\usepackage{dcolumn}
\usepackage{bm}
\usepackage{hyperref}

\begin{document}

\preprint{APS/123-QED}

\title{Collision-induced coherent dynamics}

\author{Ayanesh Maiti}
\email{ayaneshmaiti@iisc.ac.in }
 \altaffiliation{Undergraduate Department, Indian Institute of Science, Bangalore, India - 560012}
\author{Shankar Ghosh}%
 \email{sghosh@tifr.res.in}
 \affiliation{Department of Condensed Matter Physics, Tata Institute of Fundamental Research, Mumbai, India - 400005}

\begin{abstract}
In this paper we demonstrate a route to develop coherence in a system of non-driven oscillators. Here, the coherence is brought about via physical collisions through which the oscillators exchange energy. While coherence in the classical situations occurs due to sustained coupling terms in the dynamical equations, collision-induced coherence is enabled solely through strong interactions that are of intermittent nature! We demonstrate this in a Newton's cradle arrangement of oscillators by electrical and optical studies under ambient and vacuum conditions.
\end{abstract}

\maketitle

\section{Introduction}
Synchronization~\cite{Syn1,Syn2,Syn3,Syn4,Syn5,clock1,clock2,chaos1,chaos2,chaos3,chaos4,chaos5,Bio1,Bio2}, a characteristic of nonlinear dynamics, has attracted a lot of attention since its discovery in the seventeenth century. It causes two or more separate `self-sustained' oscillating objects, e.g. pendulum clocks, to exhibit collective dynamics. For this to happen, there must exist some form of sustained coupling between the oscillators and the natural frequency of each oscillator must be nearly equal. In the limit of weak coupling the individual dynamics of the oscillators are usually preserved and the effect of the coupling is mainly observed in the phase synchronisation. However, if the coupling is strong the trajectories of the oscillators tend to coincide and this leads to a complete synchronisation.

There are also instances in which a system of coupled oscillators that are not self sustained can exhibit coherent collective dynamics. In the absence of an external source of energy, these oscillators eventually come to a halt. Their collective dynamics do not fit into the description of the classical picture of synchronisation. A simple way to think of this scenario is to consider a series of pendulum clocks on a rigid (weak coupling) or a moving (strong coupling) platform whose escapement mechanisms have been defeated. The clocks are set to motion by the random impulses, but they do not receive the periodic push that otherwise keep them ticking. In time all these non-sustained clocks achieve the trivial coherent state of rest. However, in the process of doing so they exhibit interesting dynamics which in itself merits investigation.

In this paper, we present a scenario in which such non sustained oscillators can develop coherence. Unlike conventional studies, we consider a coupling between the oscillators (linear array of swinging pendulums) which is strong but intermittent in nature. Collisions between neighbouring pendulums can give rise to such strong interactions and it is through these events that the oscillators exchange energy. On the other hand, there is no interaction when the pendulums bobs are out of contact.

Collisions are a subject of fundamental interest to understand phenomena at all scales of physics. Two-body collisions are very well studied~\cite{2b_coll1,2b_coll2,2b_coll3} by tracking the energy and momentum changes or modelling the contact dynamics. However, a collision involving three or more bodies is very complex, even in a classical situation. A simple one-dimensional head-on collision between three identical spheres can have an infinite number of valid solutions~\cite{intro}! A lot of studies have worked to resolve this issue by including material deformation properties in their calculations~\cite{hz1,hz2,icp,ec1,ec2,ec3}. Some others have attempted to explain their observations by introducing the concept of a momentum transfer through pressure waves~\cite{pw}. These studies have achieved a varying degree of success in predicting the final velocities after a many-body collision, which can be measured through high-speed photography. Very few studies~\cite{ec1,ec2,ec3,ec4} have attempted to study the contact dynamics during a collision. We have generalized the framework of these experiments to demonstrate the collision-induced development of coherent dynamics in the simple one-dimensional arrangement of balls known as the Newton's cradle.

\begin{figure}
 \centering
 \includegraphics[width = 0.48\textwidth]{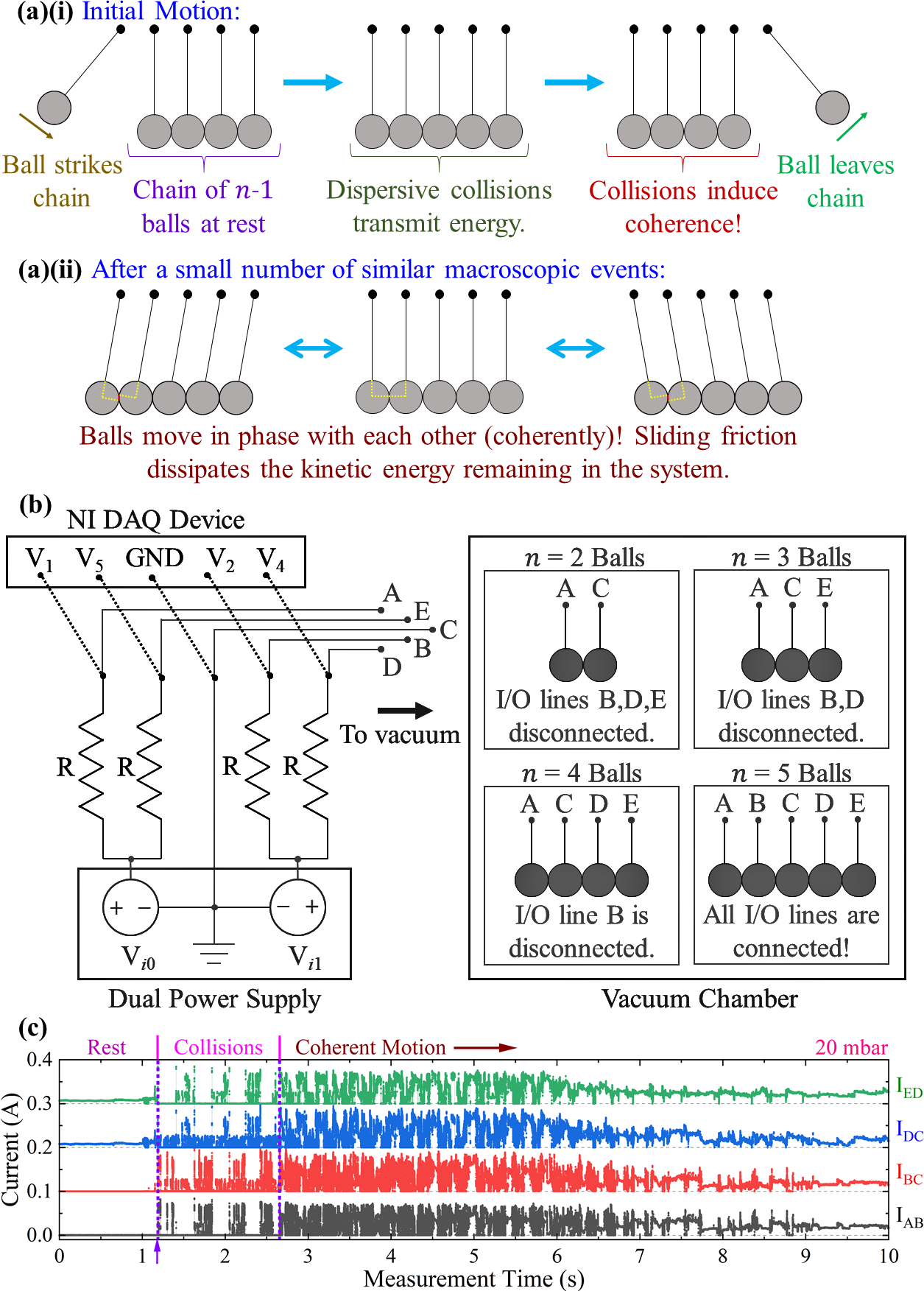}
 \caption{(a)(i) Collision-induced development of coherence and (ii) subsequent coherent oscillations if a 5-ball chain. (b) Electrical circuit for recording the contact dynamics of a Newton's cradle with 2-5 balls. (c) Measured electrical contact dynamics of a 5-ball chain in 20 mbar ambient pressure. The purple arrow marks the first collision.}
 \label{Fig1}
\end{figure}

A Newton's cradle consists of identical balls suspended using identical strings from a series of pivots on a line at a fixed height. The neighbouring pivots are separated by a distance equal to the ball diameter, so that the balls are in contact at equilibrium. When an incoming ball strikes one end of a Newton's cradle at rest, a common observation is that the striking ball comes to rest, while the ball at the other end leaves the chain with all the incoming momentum. This result, as depicted in Fig.~\ref{Fig1}(a)(i), has been widely used to demonstrate the energy and momentum conservation principles. Careful experiments~\cite{hz1} have shown that while most of the incoming momentum is indeed transferred to the ball that separates from the chain, the \textit{remaining} balls also acquire small velocities. The final velocities have been predicted to a reasonable degree of precision using various models~\cite{hz1,icp} of the one-dimensional contact interactions. These calculations show that even in an ideal situation with perfectly elastic forces, there is a significant dispersion of energy as it gets transmitted across the chain. We utilize this phenomenon to generate systems of incoherent oscillators, and find that they develop coherent dynamics as shown in Fig.~\ref{Fig1}(a)(ii). This happens even in the absence of any kind of interaction other than collisions!

\section{Experiment}
A Newton's cradle was constructed from $n$ = 2-5 stainless steel balls of 12 mm diameter suspended from a pair of acrylic rods held parallel at the same height. All of the pendulums were adjusted to an effective length $l =$~45$\pm$1 mm. The entire setup was placed in a vacuum chamber with a glass window for observation. A vertically standing steel sheet was mechanically fixed on a weak magnet placed at the bottom of the chamber at large enough distance from the balls to suppress any magnetic effects. This "arm" was mechanically coupled with a weak magnet fixed on an XY stage outside the chamber to control the balls externally. Electrical connections to an external circuit were introduced through thin copper wires that held the balls. One of the balls near the center of the chain was connected to the ground levels of a dual DC power supply and a National Instruments data acquisition (DAQ) setup. The neighbouring balls were powered to an increasing series of input voltages $V_{i0} =$~9V and $V_{i1} =$~3V through fixed Ohmic resistances $R$ = 100~$\Omega$ as shown in Fig.~\ref{Fig1}(b).

To initiate each measurement, one ball was \textit{released} from an angular displacement of about 20$^{\circ}$ away from the rest of the chain. The time evolution of the ball voltages was recorded at a sampling frequency of 10 kHz. This process was repeated 50 times for each case in ambient pressures of 20 mbar (low vacuum) and 1000 mbar (atmosphere). For consistency across our data, we have set the time to zero at the beginning of the first collision. In all of our measurements, the current $I$ flowing through the resistors was calculated using Ohm's law, $V_{\rm supplied} - V_{\rm measured} = IR$. Assuming that a negligible amount of current was drawn by the DAQ setup, this essentially gave us the current flowing to each ball in the Newton's cradle. Hence, we could easily determine the current flowing through the contacts between the balls using charge conservation principles. Fig.~\ref{Fig1}(c) shows the time evolution of the contact currents for a 5-ball chain. When two balls press into each other during a collision, the contact area between them increases along with the deformation and impact forces. This reduces the contact resistance and leads to a spike in the measured current. Thus the recorded current gives a qualitative measure of the instantaneous impact force at each contact.

Experiments were also recorded using a 60 fps camera in order to complete our description of the onset of coherence. The angular displacement $\theta$ of each ball from its mean position was tracked from these videos. The specific energy $\epsilon_i$ of the balls was calculated as:
\begin{equation}
 \epsilon_i = \mathrm{\frac{1}{2}} l^\mathrm{2} \dot{\theta}^\mathrm{2} + gl\mathrm{(1 - \cos{\theta})}
 \label{Eq1}
\end{equation}
where $g =$~9.8 m/s$^\mathrm{2}$ is the acceleration due to gravity. The first term in equation~\ref{Eq1} denotes the kinetic energy while the second term denotes the gravitational potential energy of the ball. We further calculated the total energy in the system as $\epsilon_T = \sum \epsilon_i$. For the $n =$~2 case, we further improved the time resolution of the videos by increasing the pendulum length to 60$\pm$1 mm in order to confirm some of points in the following discussion.

\section{Results and Discussion}

\begin{figure}
 \centering
 \includegraphics[width = 0.48\textwidth]{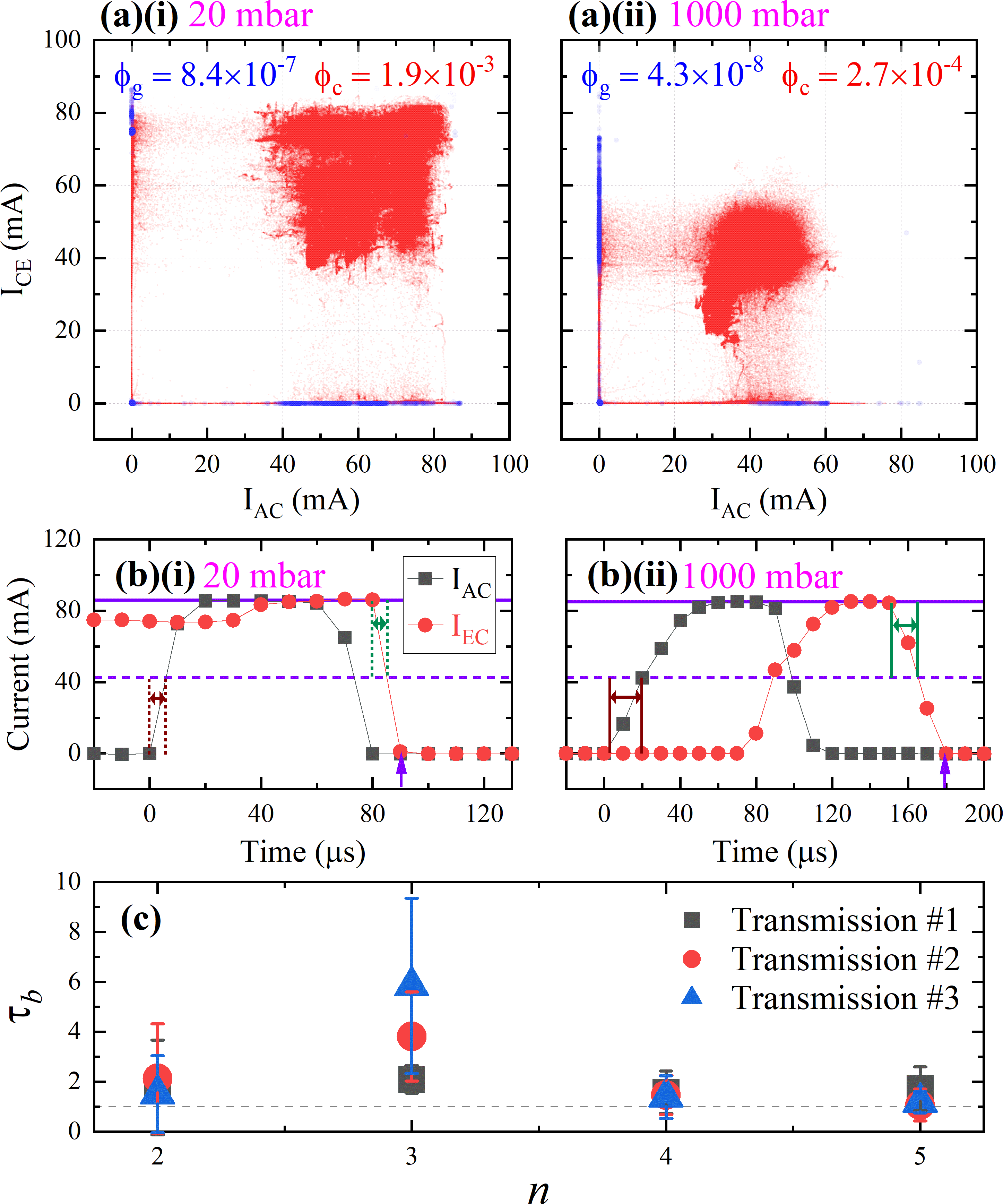}
 \caption{(a) Correlation between the contact currents $I_\mathrm{AC}$ and $I_\mathrm{CE}$ in a single 3-ball experiment performed in (i) 20 and (ii) 1000 mbar pressure during the development of coherence (blue) and the coherent motion (red). The time-averaged quantity $\phi = \langle I_\mathrm{AC}I_\mathrm{CE}\rangle$ is written in units of A$^2$. (b) Electrical contact dynamics during the first transmission event of the same experiment. The purple arrows mark the end of transmission. The violet lines indicate the maximum and half-maximum current levels. The brown and green lines show how the half-time $\tau$ of contact making and breaking is calculated. (c) Ratio of contact formation times $\tau_b=\tau_\mathrm{atm}/\tau_\mathrm{vac}$ averaged over all datasets.}
 \label{Fig2}
\end{figure}

From Fig.~\ref{Fig1}(c), we can easily identify three distinct phases of motion. Initially, all contact currents are constant. This is the part where one ball is held apart from the cradle. The entire system is at rest. When this ball is released and hits the remaining chain of balls, it generates a system of incoherent oscillators. This is clearly reflected in the observation that $I_\mathrm{AB}$ and $I_\mathrm{DE}$ take non-zero values alternately, while $I_\mathrm{BC}$ and $I_\mathrm{CD}$ vary intermittently. As this system evolves, the oscillator interact solely through collisions. A major part of the total energy gets transmitted back and forth across the chain in events similar to Fig.~\ref{Fig1}(a)(i). These \textit{transmission} events can easily be identified as the points where $I_\mathrm{AB}$ and $I_\mathrm{DE}$ alternate. After a few such events, we see that all the contact currents start showing similar variations. The system enters a coherent state of motion where all of the pendulums oscillate in-phase! This oscillation is damped mainly through the sliding friction at the contacts, as shown in Fig.~\ref{Fig1}(a)(ii).

While the rest state is trivially separated from the other observed phases, we can quantitatively distinguish between the incoherent and coherent motion by measuring the time-averaged correlation $\phi = \langle I_k I_l \rangle$ between the various contacts over short intervals. This is demonstrated through Fig.~\ref{Fig2}(a), which shows the correlation between the two contacts in a 3-ball cradle. Each symbol on the plot marks a single point of data. In the incoherent state, we observe that most of the data points are localized on the pair of straight lines $I_\mathrm{AC}I_\mathrm{CE}=$ 0. In contrast, most of the data points in the incoherent state are localized within the region with $I_\mathrm{AC}I_\mathrm{CE}>$ 0. As a consequence, the current correlation $\phi_c$ in the coherent phase is much greater than the current correlation $\phi_g$ in the incoherent phase. This observation can be understood as follows. In the coherent phase, the balls are all oscillating in-phase. They are thus constantly in contact, barring any friction-induced fluctuations. Thus all contact currents are simultaneously positive. In contrast, while the oscillations are incoherent, the balls undergo repeated collisions which lead to coherence. Balls are only in contact during the intermittent collisions, so the contact currents are zero most of the time.

One more interesting thing to note here is that in both phases of motion, the correlation $\phi$ is greater in vacuum as compared to the corresponding values in ambient pressure. It is clear easy to see from the figure that this is a direct consequence of both $I_\mathrm{AC}$ and $I_\mathrm{CE}$ values being greater in vacuum. This suggests that the contact resistance is greater in atmosphere. From this observation as well as others that are discussed later, we are able to conclude that this is due to some insulating surface impurities that are present on the balls. This layer of impurities gets removed in vacuum, leading to an increased current. The discussed results are also observed in larger chains. The difference is easiest to establish through the correlations between the contact currents at the two ends of the chain. This is because in the incoherent phase, both of these contacts can be simultaneously nonzero only during the transmission events. At all other times, either of the end balls is separated from the rest of the ball-chain! In contrast, all the correlations can be nonzero during the intermittent collisions, thus reducing the difference between $\phi_g$ and $\phi_c$ in the corresponding cases. Now that we have sufficiently resolved the coherent phase of motion from the preceding incoherent phase, we turn our focus to the collisions that lead to the development of coherence.

\begin{figure}
 \centering
 \includegraphics[width = 0.48\textwidth]{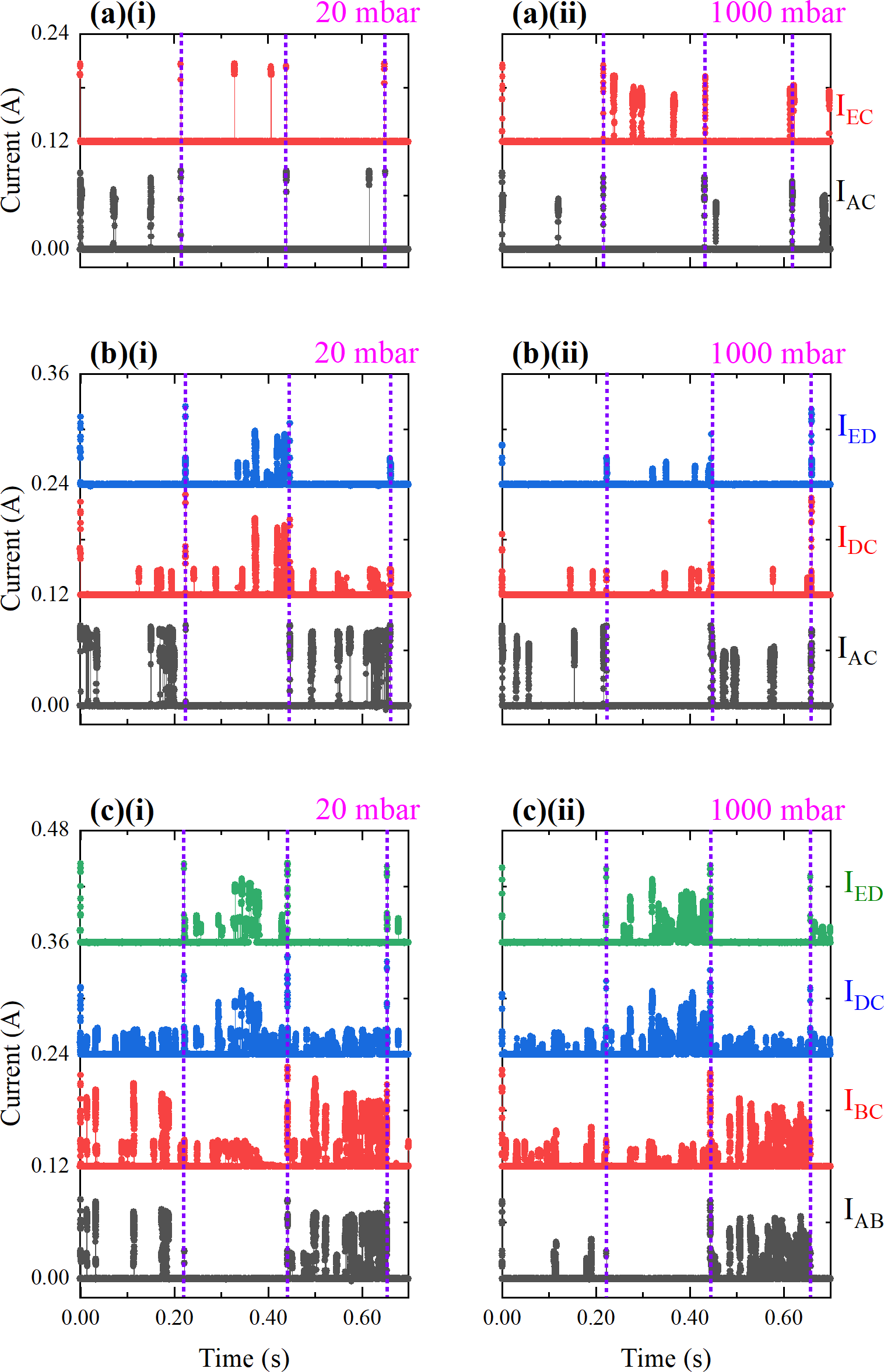}
 \caption{Electrical contact dynamics following the first transmission event of (a) 3, (b) 4 and (c) 5 ball chains in (i) 20 and (ii) 1000 mbar ambient pressure. The current data is successively offset by 0.12 A for clarity. The purple lines mark transmission events.}
 \label{Fig3}
\end{figure}

The electrical contact dynamics during the first transmission event in a 3-ball chain is expanded in Fig.~\ref{Fig2}(b). We observe an ordered sequence of collisions as the incident energy passes through the chain. Similar sequences are observed in all of our experiments for all chain lengths. This is in agreement with the discussed literature. Notably, there is a strong dependence of the initial contact current $I_\mathrm{EC}$ on the pressure. This is due to the removal of insulating surface impurities in a vacuum, as mentioned earlier. One striking new feature is found in the pressure dependence of the current profiles. The contact forms and gets broken much more rapidly in vacuum than in atmospheric pressure, i.e., $\tau_b=\tau_\mathrm{atm}/\tau_\mathrm{vac} \gg 1$. Here $\tau_\mathrm{atm}$ and $\tau_\mathrm{vac}$ are the half time associated with the formation of the contact in ambient and vacuum conditions, respectively. Fig.~\ref{Fig3}(b) shows that the half-time of these contact formation processes change drastically from less than 10 $\mu$s in 20 mbar to around 15-20 $\mu$s in 1000 mbar, i.e, $\tau_b\approx$ 2 during the first collision. For n = 3 balls, this number is also found to increase in the following transmissions, as plotted in Fig.~\ref{Fig3}(c). This again supports the existence of the insulating surface layer in ambient pressure. The contact establishment in atmospheric pressure would involve the rupturing of these layers, thus taking slightly more time than the breaking of contacts. The longer total transmission time also supports these arguments. Interestingly, the contact time does not show any dependence on pressure for cradles that contained more than three balls. This indicates that the insulating layer is not broken in these cases. This is consistent with the intuitive expectation that a longer chain of balls should have dampened impacts at the individual contacts.

As shown earlier in Fig.~\ref{Fig1}(c), the balls that are left in the chain have a nontrivial contact dynamics between transmissions. This is expanded for one set of measurements in Fig.~\ref{Fig3}. It is clear that these $n$-1 balls undergo intermittent collisions that produce sharp pulses in the measured contact current. These events imply the fragmentation of the chain due to the dispersion of energy during the first transmission event. These results are also in agreement with the literature discussed in the introduction. Notably, the intermittent collisions between these balls seem to be quite random. The time sequence of contact pulses shows large variations from one measurement to another. It is evident that while the oscillators corresponding to the remaining balls have finite energies and phases, small differences in some situational parameters cause the exact distributions to show large fluctuations. This makes the system ideal for demonstrating the collision-induced development of coherence for varied initial conditions.

Despite the large fluctuations involved, we intuitively expect to see one particular trend. As the number of balls in the chain is increased, there should be a clear rise in the frequency of the intermittent collisions. Firstly, for $n$ = 2, no collisions can occur in a chain with just $n$-1 = 1 ball. For $n$ = 3, the two balls that are left in the chain are able to collide. However, once a collision ends, the balls move in separate directions and cannot undergo a second collision for a considerable duration of time! In contrast, for a chain with $n$ = 4 balls, there are $n$-1 = 3 balls left behind. Once two balls separate after a collision, the third ball can collide with one of these balls and reflect it back. This can consequently initiate a secondary collision between the initially colliding balls after arbitrarily short delays! For the $n$ = 5 ball case there can be similar but even more complicated situations allowing for more frequent collisions. In fact, we observe these exact trends in Fig.~\ref{Fig3}. While the intermittent collisions spikes are clearly distinct in the 3 ball case, they appear to become arbitrarily close for 4 and 5 balls.

\begin{figure}
 \centering
 \includegraphics[width = 0.48\textwidth]{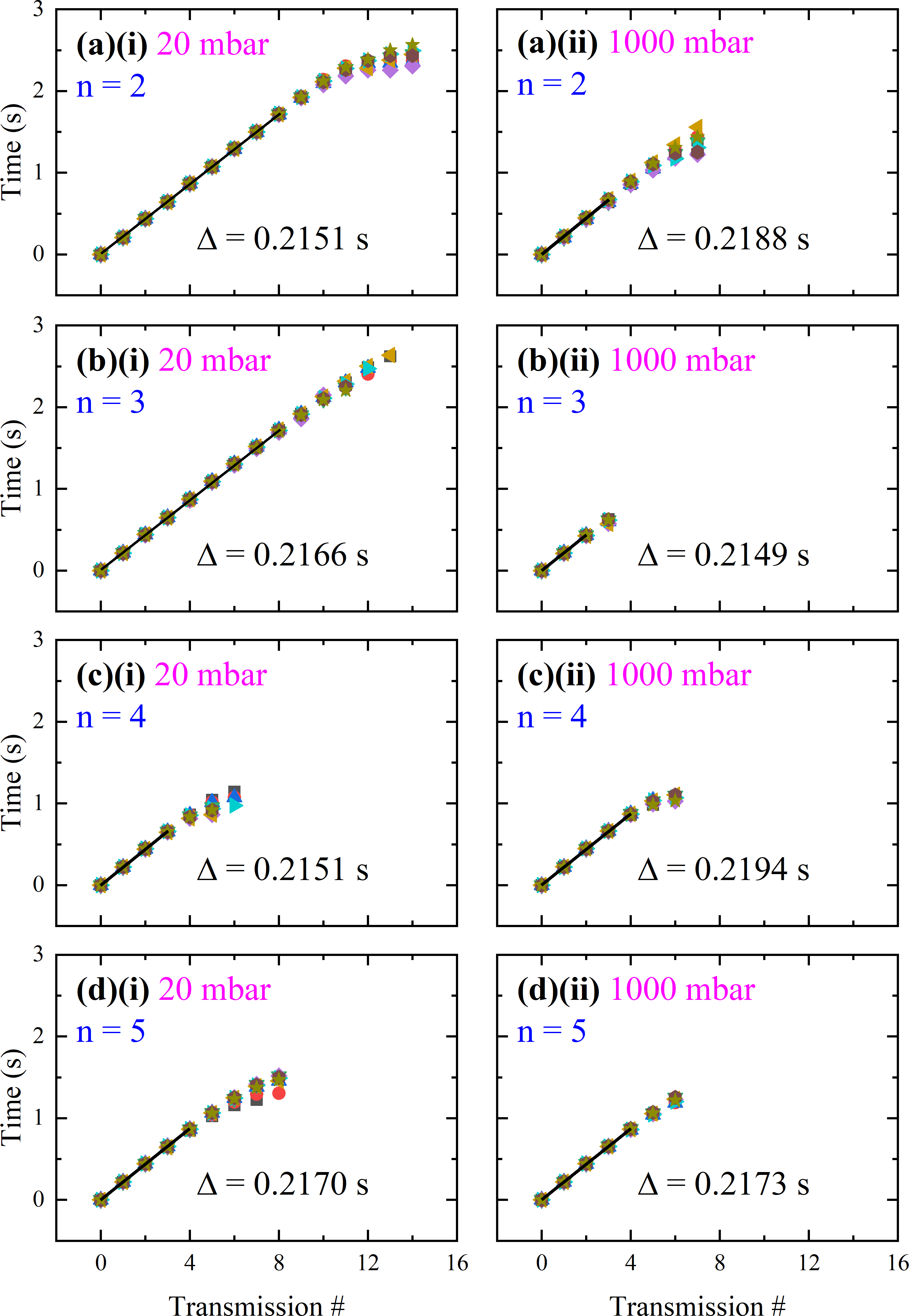}
 \caption{Starting time of distinguishable transmission events observed in nine different measurements with (a) 2, (b) 3, (c) 4 and (d) 5-ball chains in (i) 20 and (ii) 1000 mbar ambient pressure. Collective fits of the initial linear part (black line) have slope $\Delta$ per collision.}
 \label{Fig4}
\end{figure}

In all cases, none of the observed collisions is perfectly elastic. In fact, each collision is slightly inelastic and serves to dissipate a slight amount of energy. This effectively acts like a noisy and impulsive damping force which is what causes the development of coherence. In all of our experiments, we physically observe that the balls end up oscillating coherently after a short period of collisions (around 1-4 seconds). The coherent oscillations are mainly damped by the contact friction. This damping is quite weak in comparison to the impulsive collision forces that are responsible for the growth of coherence. As a result, the system continues to perform underdamped collective oscillations for a very long duration (around 40-120 seconds depending on the length of the chain). Fig.~\ref{Fig4} shows the time sequence of transmissions in nine different experiments. Except for the first transmission which produces a random configuration, all of the others occur during the coherence growth.

There is one trend in Fig.~\ref{Fig4} that appears quite unusual at the first glance. While we seem to observe no pressure dependence for $n\geq$ 4, the case is drastically different for $n\leq$ 3. In fact, the number of transmissions for a 3 ball chain increases about 4$\times$ when the pressure is reduced from 1000 mbar to 20 mbar! For $n$ = 2, the increase is about 2$\times$. There is a proportional increase in the duration of coherence growth. However, this behaviour can be understood as follows. Since collisions are relatively infrequent for $n\leq$ 3, they tend to occur with large differences in initial velocities. In atmospheric pressure, the surface impurity layers contribute to an additional loss of energy as they get disrupted during collisions. In vacuum these layers are removed and hence less energy is lost per collision. Since the energy dissipation is required for the development of coherent dynamics, it is evident that the balls would synchronize much faster in atmospheric pressure. For $n\geq$ 4, the intermittent collisions occur frequently and with much lower velocity differences. Hence the surface layers do not get significantly disrupted! Thus there is no clear pressure dependence. Also, since the damping strength is intermediate, the number of transmissions for $n\geq$ 4 is expected to be between the high and low pressure limits of the $n\leq$ 3 cases. This is indeed what we observe! The results are also consistent with our previous discussion.

\begin{figure}
 \centering
 \includegraphics[width = 0.48\textwidth]{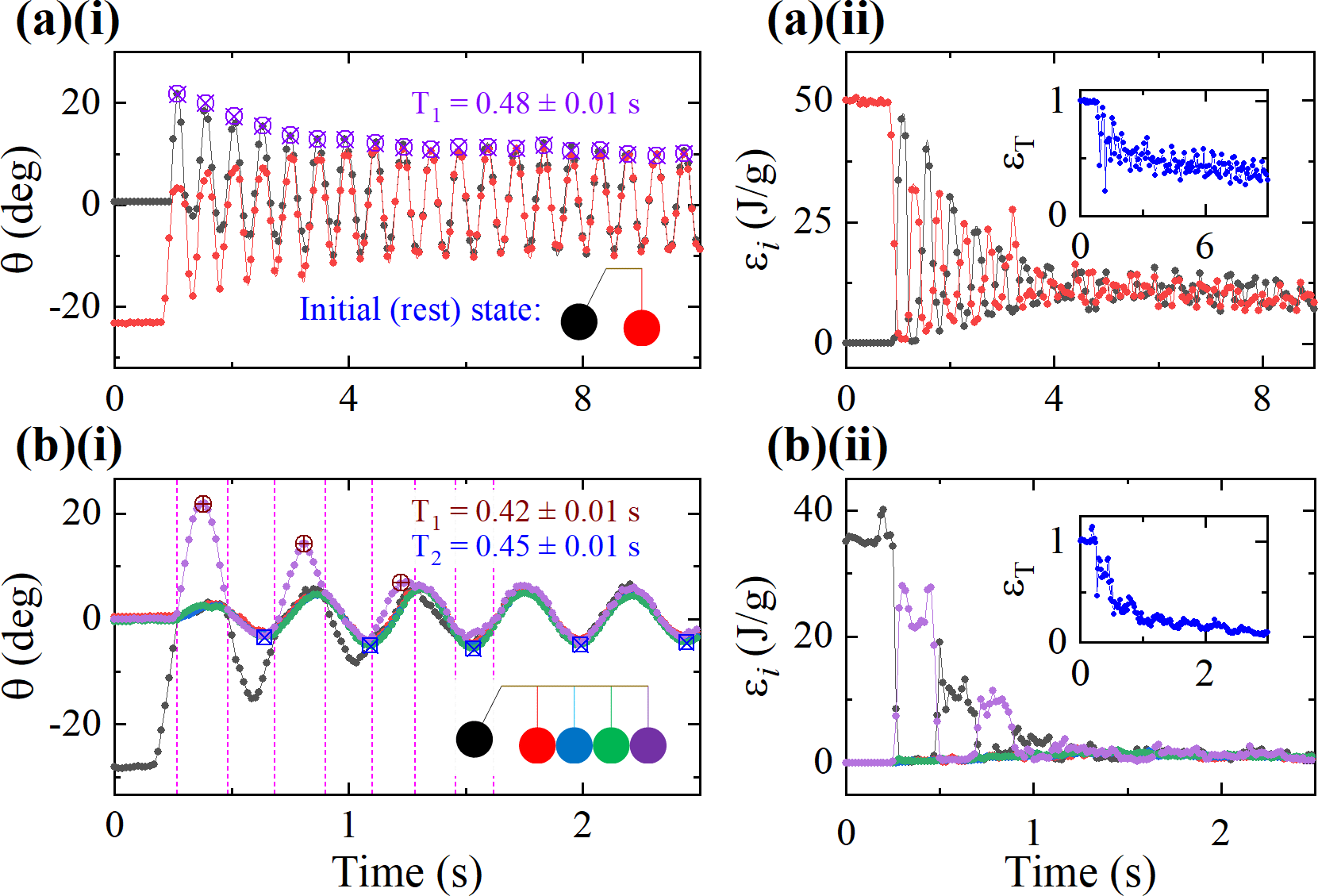}
 \caption{Time evolution of (i) the angular displacement $\theta$ and (ii) the specific energy $\epsilon$ of the balls in the (a) 2 and (b) 5 ball chains in 20 mbar vacuum. The insets in (ii) show the total energy $\epsilon_T$ of the system in units of its initial value. The pink lines identify the transmission events. The oscillation periods of the isolated ball $T_1$ and the remaining chain $T_2$ are estimated from the marked points.}
 \label{Fig5}
\end{figure}

Another notable thing that we observe in Fig.~\ref{Fig4} is that initially the transmissions are linearly spaced in time. The separation $\Delta$ between consecutive collisions, calculated from the slope of the black line in Fig.~\ref{Fig4}, is about half the period of oscillation ($\approx$ 0.21 seconds). The process deviates from this linearity as the coherence increases. To explain this, we examine the data from our videos. Fig.~\ref{Fig5} shows the growth of coherence of the balls observed in one $n$ = 2 and one $n$ = 5 experiment, both performed in a 20 mbar vacuum. We observe an especially fascinating feature in the 5 ball case. The \textit{remaining} balls seem to oscillate at a different frequency compared to the \textit{isolated} one. In fact, while the isolated ball evolves exactly like a simple pendulum, the other balls act effectively like a damped oscillator\footnote{For a damping force $F_d$ proportional to the velocity $v$ so that $F_d = -bv$, a damped harmonic oscillator of mass $m$ moves with a reduced angular frequency $\omega_d = \sqrt{\omega_n^2 - \gamma^2}$ where $\gamma = b/$2$m$ }! The most prominent source of damping present are the intermittent collisions. This collision-induced damping reduces the angular frequency $\omega$ = 2$\pi/T$ from about 14.96 rad/s to about 13.96 rad/s. This causes a phase mismatch to develop between the isolated ball and the rest of the chain. As a result, transmissions do not occur near the equilibrium configurations. In fact, we observe that as the system evolves, transmissions start increasingly closer to the extreme positions. As this happens, they become more and more frequent and the coherence increases. Hence we complete our description of the collision-induced development of coherence.

There are a few remaining points that are evident from Fig.~\ref{Fig5}. Firstly, there is no change in the oscillation frequency of the $n$ = 2 case, even after the system seems synchronized. This shows that the frictional damping is indeed much weaker than the collisional damping that induces the observed coherent dynamics. Also, if we look at the specific energies, we observe that there is much more dissipation in the $n$ = 5 case as compared to the $n$ = 2 case. This is due to the lack of intermittent collisions in the latter case. From these results, we conclude that the collisional dissipation acts as the limiting factor as we increase the number of oscillators and the dimensionality of their spatial distribution.

\section{Conclusions}
In conclusion, we have introduced a new mechanism of coherence development that involves collisions. In contrast to the previously researched systems of weakly interacting self-sustained oscillators, our study establishes the possibility of coherence in systems that only interact through intermittent impulsive forces. As long as there is a mechanism to generate frequent collisions in the system, the route to coherence via intermittent collision hold true for both small and large populations of oscillators,

\bibliography{ref}

\end{document}